# Linear Magnetoresistance in Topological Insulator Thin Films: Quantum Phase Coherence Effects at High Temperatures

B. A. Assaf,[1] T. Cardinal,[1] P. Wei,[2] F. Katmis,[2,3] J.S. Moodera[2,3] and D. Heiman[1]

[1]Department of Physics, Northeastern University, Boston, MA 02115
[2]Francis Bitter Magnet Lab, Massachusetts Institute of Technology, Cambridge, MA 02139
[3]Department of Physics, Massachusetts Institute of Technology, Cambridge, MA 02139

*In addition to the weak antilocalization cusp observed in the magnetoresistance (MR) of topological insulators at low temperatures and low magnetic fields, we find that the high-field MR in $Bi_2Te_2Se$ is linear in field. At fields up to B=14T the slope of this linear-like MR is nearly independent of temperature over the range T=7 to 150K. We find that the linear MR arises from the competition between a logarithmic phase coherence component and a quadratic component. The quantum phase coherence dominates up to high temperatures, where the coherence length remains longer than the mean free path of electrons.*

Three-dimensional (3D) topological insulators have been predicted to host metallic surface states surrounding an insulating bulk as a result of a band inversion in compounds containing elements having a large spin-orbit interaction.[1,2] These electron states have been shown to exhibit strong spin-momentum locking that is attractive for potential applications in spintronic devices. $Bi_2Se_3$ and $Bi_2Te_3$ have been experimentally confirmed to have these topological properties through the observation of the surface Dirac dispersion using angle-resolved photoelectron spectroscopy (ARPES).[3]

Considerable effort has thus been focused on probing the surface states with magnetotransport experiments. Quantum coherence effects,[4,5,6,7] quantum oscillations[8,9,10], as well as some non-linear Hall signatures[9,10] have been observed and well-studied in many cases. An interesting observation was made by several groups recently; the magnetoresistance (MR) shows a non-saturating positive, linear-like trend at high fields[7,8,9,11]. It has been suggested[7,8] that this linear-like MR response arises from the linear Dirac surface dispersion, which according to the quantum theory of linear MR (LMR) developed by Abrikosov should result in such a phenomenon.[12] Some studies have also suggested that the LMR behavior becomes suppressed as the bulk is depleted of carriers.[11] This raises the question of the origin of the LMR.

We report on magnetotransport measurements of $Bi_2Te_2Se$ thin films grown by molecular beam epitaxy (MBE). At low magnetic fields the MR exhibits a weak antilocalization (WAL) cusp that becomes suppressed when the temperature is increased. In addition, at high fields the MR increases with increasing field in a nonsaturating, linear trend up to B = 14 T. We account for MR(B) over the entire range of fields and temperatures with a *modified* Hikami, Larkin and Nagaoka (HLN) quantum interference model.[13] It simultaneously accounts for the quantum phase interference cusp at low fields as well as the linear-like MR at high fields. It is shown that an additional quadratic term - consisting of both quantum and classical components - compensates the logarithmic dependence of the quantum interference at high fields, leading to an intermediate linear-like MR field dependence. However, we show that even at high temperatures the quantum interference dominates, indicating that electrons retain quantum phase coherence at high temperatures, possibly up to room temperature.

Thin films of $Bi_2Te_2Se$ were grown on Si (111) at 200 °C using MBE. The base pressure in the MBE chamber was ~ $7\times10^{-10}$ torr prior to the growth. Flux monitors in the MBE system allow the evaporated thickness to be monitored in real time during the growth.[14] The stochiometry was calibrated prior to the growth and monitored during the growth to achieve a 2:2:1 ratio of Bi:Te:Se. This ratio was later confirmed by energy dispersive X-ray spectrometry. X-ray diffraction data showed good *c*-axis alignment for all films, having (000$\ell$) diffraction peaks up to $\ell$ = 21 as shown in Fig. 1. We confirmed the formation of a single phase and found no evidence of peak splitting. The inset of Fig. 1 shows the (0006) reflection peak with Kiessig fringes in the vicinity of the Bragg peak corresponding to the 15 nm thickness of the film. Magnetotransport measurements were performed on photolithography-patterned Hall bars (1 mm long and 0.3 mm wide). The samples were measured at temperatures down to T = 7 K and in magnetic fields up to $B$ = 14 T. The Hall voltage was found to be linear in field and the electron density was ~ $5\times10^{19}$ $cm^{-3}$.

Although MBE-grown Bi$_2$Te$_2$Se has not been previously reported, it is expected to be the most insulating among the Bi-based topological insulator materials as it lies between n-type Bi$_2$Se$_3$ and p-type Bi$_2$Te$_3$ when grown in bulk. Also, the ordering of the Se and Te atomic layers presumably inhibits Se from vacating the lattice.[10,15] Unfortunately, the present MBE-grown films have not yet exhibited strong insulating characteristics possibly due to the fact that excess Se was not used during the growth. The present sample quality is similar to that previously reported for Bi$_2$Te$_2$Se nanoplatelets.[16]

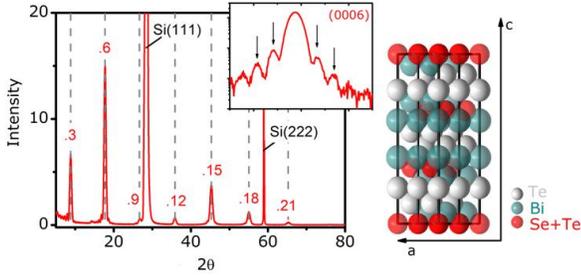

Figure 1. (Color Online) θ-2θ X-ray diffraction spectra of a 15 nm thick Bi$_2$Te$_2$Se film on Si acquired using Cu K$_\alpha$ radiation. The gray dashed lines correspond to the calculated diffraction peaks of the hexagonal structure with a = 0.431 nm and c = 3.001 nm, labeled by their hexagonal axis indices. The inset shows an enlarged view of the (0006) reflection peak to clarify the Kissieg fringes in the vicinity of the Bragg peak. The crystal unit cell is shown on the right.

The results of MR($B$) measurements for a 15 nm thick Bi$_2$Te$_2$Se film are shown in Fig. 2 for fields up to 14 T and temperatures from T = 7 to 147 K. A well-pronounced WAL cusp is observed at low temperatures in the low-field region. As the temperature is increased, the cusp weakens as a result of the decreasing coherence length and eventually gives rise to a quadratic-like $B$-dependence at low fields. At higher fields the MR does not saturate and follows a linear-like dependence. In this linear region the slope, $d$MR/$dB$, changes very little with temperature, only changing by about 20 % over the large range of temperatures from T = 7 to 147 K. Below, we show that by including additional higher order scattering terms in the HLN model and a classical cyclotronic contribution, the MR of a topological insulator can be fit at all fields and all temperatures.

The low-field MR of topological insulators is well-described by the Hikami-Larkin-Nagaoka (HLN) quantum interference model.[13] The quantum correction to the magnetoconductance in 2D systems is given by

$$\Delta G(B) = -\frac{e^2}{2\pi h}\left[\psi\left(\frac{B_\phi}{B}+\frac{1}{2}\right) - \ln\left(\frac{B_\phi}{B}\right)\right]$$
$$-\frac{e^2}{\pi h}\left[\psi\left(\frac{B_{SO}+B_e}{B}+\frac{1}{2}\right) - \ln\left(\frac{B_{SO}+B_e}{B}\right)\right]$$
$$+\frac{3e^2}{2\pi h}\left[\psi\left(\frac{(4/3)B_{SO}+B_\phi}{B}+\frac{1}{2}\right) - \ln\left(\frac{(4/3)B_{SO}+B_\phi}{B}\right)\right]. \quad (1)$$

Here, $B_i$ are the characteristic fields of each respective scattering channel ($i = \phi$, $SO$, $e$) given by $B_i = \hbar/(4eL_i^2)$. $L_\phi$ is the phase coherence length, $L_{SO}$ is the spin-orbit scattering length and $L_e$ is the elastic scattering length (or the mean free path). For small fields the phase coherence scattering term (first term) dominates since the coherence length is the longest of the three lengths. The spin-orbit and elastic scattering lengths yield characteristic fields of the order of several tesla. A maximum field of 14 T would therefore allow us to cover the low-field regime of the two latter terms in Eq. (1). It can be shown that the latter two terms containing spin-orbit scattering and elastic scattering can thus be approximated by a $B^2$ parabola.[13] These approximations lead to

$$\Delta G(B) = -\frac{\alpha e^2}{\pi h}\left[\psi\left(\frac{\hbar}{4eL^2B}+\frac{1}{2}\right) - \ln\left(\frac{\hbar}{4eL^2B}\right)\right] + \beta B^2. \quad (2)$$

α is included to account for 2D bulk effects and possible contributions from the top and bottom surfaces of the film. $L$ is the phase coherence length and β is the quadratic coefficient arising from the additional scattering terms. The first term in Eqs (1) and (2) will be referred to as the *simple* HLN model. One must also keep in mind that conventional cyclotronic MR might have to be taken into account. β would therefore be composed of a classical cyclotronic component $\beta_c$ in addition to the quantum scattering terms arising from the last two terms in the HLN model that will be referred to as $\beta_q$.

There is, however, a requirement that must be met in using this model at high temperatures. Quantum interference requires that the phase coherence length be longer than the elastic mean free path ($\ell$) of the scattering electrons. In the present samples the phase coherence length is about an order of magnitude larger than the elastic mean free path at low temperatures. At T = 7 K, $L$ = 58 nm and $\ell$ = 3.9 nm, while at T = 147 K, $L$ = 9.4 nm and $\ell$ = 3.5 nm, where $\ell$ is obtained from the measured Hall mobilities and the Fermi wavevector $k_F$ = 0.060 Å$^{-1}$ measured in Bi$_2$Te$_2$Se. The ratio $L/\ell$ is thus 15 at low temperatures and at least 2.7 at high temperatures. When $L/\ell$ is large the electrons make enough elastic collisions before losing phase coherence. We fit $\Delta G(B)$ to the experimental data for the entire range of fields up to $B = 14$ T. The fitted curves from Eq. (2) and the data for $\Delta G(B)$ are shown in Fig. 3 for the $Bi_2Te_2Se$ film at T = 7 and 147 K. There is excellent agreement over the entire field range. Similar fits were also carried over to the inverted MR($B$) data and plotted as solid curves in Fig. 2. *Equation (2) was found to describe the experimental data at all measured fields and temperatures.*

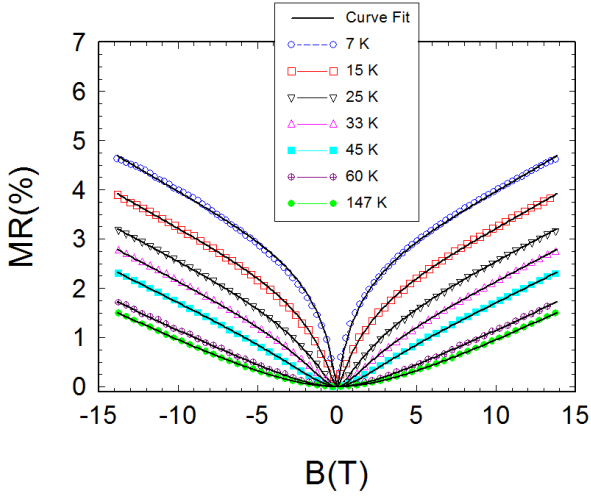

Figure 2. (Color online) Magnetoresistance versus applied magnetic field for a 15 nm thick $Bi_2Te_2Se$ film at temperatures between T = 7 K and 147 K. The solid curves represent the modified HLN model fit to the data at all temperatures and all fields.

Furthermore, in order to separate out the quantum phase-coherent contribution to the MR, Fig. 3 plots the field-dependence of the two terms of Eq. (2) separately. Note that the decreasing slope ($d\Delta G/dB$) of the phase-coherent term is compensated by the increasing slope of the quadratic term, leading to a constant slope. The phase-coherent term in the simple HLN model is seen to deviate from the data beginning at around 4 T. Below 4 T, fitting the data to either the simple or modified HLN model yields similar fitting parameters. This indicates that the modified model accurately accounts for the WAL cusp. The LMR on the other hand is seen to arise at high fields as an intermediate regime where the quadratic term compensates the saturating logarithm in the simple HLN model. Thus, the quadratic term makes its most significant contribution in the high field regime. It is interesting that the quantum coherent contribution to the MR is clearly much larger than the quadratic contribution at temperatures as high as 147 K. This reveals that electrons are able to retain phase information at high temperature, which can be technologically relevant if it is shown to hold at room temperature.

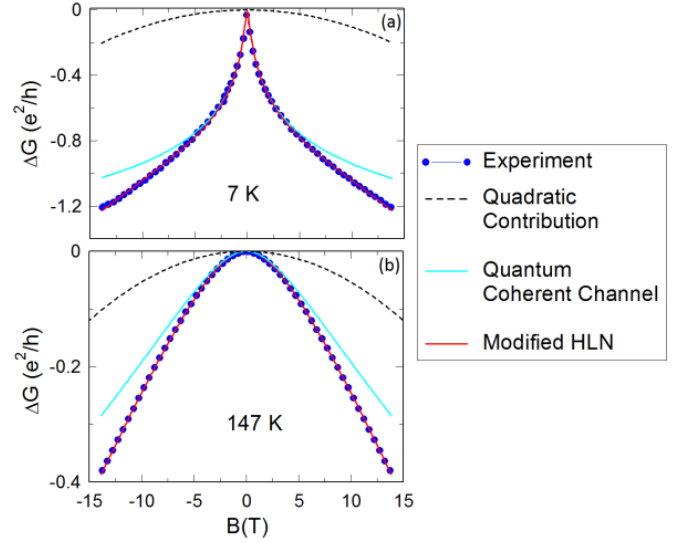

Figure 3. (Color online) Magnetoconductance of $Bi_2Te_2Se$ at (a) 7K and (b) 147 K, showing individual contributions from the phase-coherent term and the quadratic term in the modified HLN model. The quantum phase-coherent contribution is clearly dominant even at high temperatures.

The fitting parameters $\alpha(T)$ and $L(T)$ were extracted from the $Bi_2Te_2Se$ MR data and plotted in Fig. 4. At T = 7 K we find $\alpha$ = 0.85, which matches previous reports on WAL in topological insulators.[4,5,6] An increase in $\alpha$ is observed with increasing temperature. This has been attributed to a coherent surface-to-bulk scattering channel in the regime where the surface-to-bulk scattering length is smaller than the coherence length.[5] The behavior of the phase coherence length $L(T)$ shows a decaying $T^{-\gamma}$ power law trend, with $\gamma = 0.75 \pm 0.05$. This temperature dependence is complicated and can have contributions from several effects. At low temperatures a decay can arise from inelastic electron-electron collisions, which would result in $\gamma = 0.5$ for 2D, and $\gamma = 0.75$ for 3D.[17,18] 18However, electron-phonon interactions can have the same effect when the temperature is higher, resulting in a faster decay with $\gamma = 1$.[18] It is difficult to attribute the decay to a specific mechanism since our measurements span a large temperature range where various contributions result in different temperature variations.

Next, we consider the origin of the quadratic component $\beta$. As discussed previously, spin-orbit and elastic terms in the HLN model as well as cyclotronic MR all have to be taken into account, as $\beta = \beta_q + \beta_c$. The quantum part of $\beta$ can be approximated by the following:[13]

$$\beta_q B^2 = -\frac{e^2}{24\pi h}\left[\frac{B}{B_{SO}+B_e}\right]^2 + \frac{3e^2}{48\pi h}\left[\frac{B}{(4/3)B_{SO}+B_\phi}\right]^2 \quad (3)$$

Since $L_{SO} > L_e$, $\beta_q$ provides a positive quantum correction to $\Delta G$. However, the experimental data requires a total negative correction to $\Delta G$. We thus need to incorporate a conventional negative cyclotronic term in $\Delta G$ of classical origin given by

$$\beta_c B^2 = -\mu_{MR}^2 G_o B^2 \qquad (4)$$

where $\mu_{MR}$ is the MR mobility[19] and $G_0$ is the zero-field conductance. We extract the value $L_{SO} = 6$ nm at T = 7 K from $\beta = -1.1 \times 10^{-8}$ $\Omega^{-1}T^{-2}$ and the assumption that $\mu_{MR} \approx \mu_H$.[19] This is a reasonable estimate for $L_{SO}$ compared to values reported in the literature for $Bi_2Se_3$.[5]

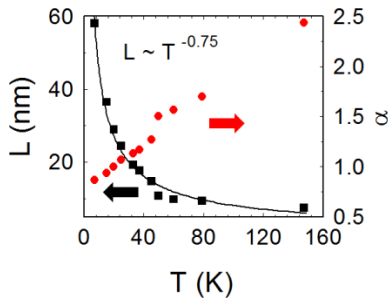

Figure 4. (Color Online) Phase coherence length $L(T)$ and $\alpha(T)$ plotted versus temperature for the $Bi_2Te_2Se$ film.

Apart from this model, there are two common models used to explain linear MR, the classical LMR model of Parish-Littlewood[20,21,22] and the quantum LMR model of Abrikosov.[12] In the classical model, the slope of the LMR, is expected to vary with mobility as the temperature changes. This has been shown experimentally in films containing random MnAs nanoparticles in a GaAs matrix, where the mobility is linearly proportional to $dMR/dB$ as they both vary with temperature by more than an order of magnitude.[22] For $Bi_2Te_2Se$, the Hall mobility decreases by only 7 % over the range of temperatures, while the slope $dMR/dB$ decreases by 20 %. Although both quantities decrease with increasing temperature, they lack a quantitative correlation. Alternatively, in the quantum LMR model, it is assumed that the magnetic field is high enough to reach the magnetic quantum limit were electrons occupy only the lowest Landau level.[12] Here, our Fermi level is too high for such a quantization to occur. Also, we do not observe any quantum oscillations up to 14 T, hence ruling out the possibility of finite Landau level filling. However, at extremely high magnetic fields or in samples having a much lower carrier density the magnetic quantum limit may be reached and quantum LMR may be applicable. In any case, as $B$ approaches $1/\mu$, Shubnikov de-Haas oscillations should arise and mask the LMR.

In conclusion, we have shown that a *modified* Hikami-Larkin-Nagaoka model of quantum phase coherence can describe the magnetic-field-dependent MR of the topological insulator $Bi_2Te_2Se$ at all magnetic fields and a wide range of temperatures. At low fields the simple HLN model provides an excellent fit to the cusp in MR($B$). However, at high fields the MR($B$) data deviates from the simple HLN model and this deviation requires an additional quadratic contribution. The sum of these two competing components (logarithmic and quadratic) then leads to a linear-like MR up to at least fields of 14 T. We thus find that the MR in $Bi_2Te_2Se$ and several other systems is linear in field because of compensation between quantum and classical contributions, which arise from higher-order terms in the HLN model and from cyclotronic MR. It is remarkable that coherent quantum interference prevails at temperatures as high as T ~ 150 K. In principle, quantum phase coherence can be dominant in thin disordered samples where the mean free path is short, but the coherence length could still be long enough to contribute even at room temperature.

Note added in proof: Recent measurements of the MR in $Bi_2Te_3$ have shown a linear-like trend up to 60 T.[23] The full HLN model (our Eq. (1)) yielded a good fit to that data. This agrees precisely with our assertion that the spin-orbit and elastic terms in the full HLN model yield a significant contribution to the parabolic term. Terms with characteristic fields of the order of the maximum applied field will produce a parabolic functional form (Eq. (3)) that compensates the phase coherence of the simple HLN formula at high fields. This is evidently the case with their term containing $B_{SO}$ and $B_e$. It is unclear, however, how much classical MR contributes to the compensation in their case, nevertheless, the fact that $L_{SO} < L_e$ causes the contributions from $\beta_q$ and $\beta_c$ to have the same negative sign.

Acknowledgements
This work was supported by the grant DMR-0907007 from the National Science Foundation and in part by the MRSEC Program of the National Science Foundation under award number DMR-0819762. JSM is also partly supported by NSF DMR-0504158 and ONR-N00014-09-1-0177 grants. We thank H. Steinberg, G. Creeth and R. Markiewicz for useful conversations.